\documentstyle[emulateapj]{article}

\slugcomment{\it Submitted to ApJ January 12, 2001}

\begin{document}


\title{Detection of Cosmic Shear with the {\it HST} Survey Strip}

\author{Jason Rhodes$^{1,2}$, Alexandre Refregier$^{3}$ \& Edward
J. Groth$^{2}$} 

\affil{1 Code 681, Goddard Space Flight Center, Greenbelt, MD 20771; jrhodes@band1.gsfc.nasa.gov}
\affil{2 Physics Department, Princeton University,
Jadwin Hall, P.O. Box 708, Princeton, NJ 08544;
groth@pupgg.princeton.edu} 
\affil{3 Institute of Astronomy, Madingley
Road, Cambridge, CB3 OHA, U.K.; ar@ast.cam.ac.uk}

\begin{abstract}
Weak lensing by large-scale structure provides a unique method to
directly measure matter fluctuations in the universe, and has recently
been detected from the ground. Here, we report the first detection of
this `cosmic shear' based on space-based images. The detection was derived from
the Hubble Space Telescope (HST) Survey Strip (or ``Groth Strip''), a
$4' \times 42'$ set of 28 contiguous WFPC2 pointings with $I <
27$. The small size of the HST Point-Spread Function (PSF) affords
both a lower statistical noise, and a much weaker sensitivity to
systematic effects, a crucial limiting factor of cosmic shear
measurements.  Our method and treatment of systematic effects were
discussed in an earlier paper (Rhodes, Refregier \& Groth 2000).  We
measure an  rms shear of 1.8\% on the WFPC2 chip scale (1.27'),
in agreement with the predictions of cluster-normalized CDM
models. Using a Maximum Likelihood (ML) analysis, we show that our
detection is significant at the 99.5\% confidence level (CL), and
measure the normalization of the matter power spectrum to be
$\sigma_{8} \Omega_{m}^{0.48} = 0.51^{+0.14}_{-0.17}$, in a
$\Lambda$CDM universe.  These 68\% CL errors include (Gaussian) cosmic
variance, systematic effects and the uncertainty in the redshift
distribution of the background galaxies. Our result is consistent with
earlier lensing measurements from the ground, and with the
normalization derived from cluster abundance. We discuss how our
measurement can be improved with the analysis of a large number of
independent WFPC2 fields.
\end{abstract}

\section{Introduction}
\label{introduction}
Weak gravitational lensing has now been established as a unique
technique to measure the mass distribution in the universe (see
Mellier 1999\markcite{mel99}; \markcite{bar00}Bartelmann \& Schneider
2000 for reviews). Recently, several groups detected this effect in
the field by measuring the coherent distortions it induces in the
shape of background galaxies (\markcite{wit00}Wittman et al. 2000;
\markcite{wae00}van Waerbeke et al.  2000; \markcite{bac00a}Bacon,
Refregier \& Ellis 2000a (BRE); \markcite{kai00}Kaiser, Wilson \&
Luppino 2000; \markcite{mao00} Maoli et al. 2000). This opens wide
prospects for the measurement of the distribution of dark matter on
cosmological scales and of cosmological parameters.

In this paper, we present the first detection of this `cosmic shear'
based on space-based images. For this purpose, we use the HST Survey
Strip (or ``Groth Strip''), a $4'\times 42'$ survey consisting of 28
contiguous WFCP2 fields with a limiting magnitude of $I \approx 27$
(Groth et al. 1994).  A shear measurement method adapted to HST
images and a detailed study of systematic effects for the Strip was
presented in an earlier paper \markcite{rho99}Rhodes, Refregier \&
Groth (2000, RRG; see also \markcite{rhot99}Rhodes 1999). The main
advantage of HST images for weak lensing is the small PSF (0.1''
compared to $\sim 1''$ from the ground), which allows a larger surface
density of resolved galaxies and makes the shear measurement
much less sensitive to the PSF smearing.  We apply the results to the
Strip and search for a lensing signal using a Maximum Likelihood (ML)
approach. We then derive constraints on the amplitude of the mass
power spectrum and compare it to other measurements.

The paper is organized as follows. In \S\ref{theory} we summarize the
main theoretical results which serve as a reference for our
measurement. In \S\ref{data} we describe the relevant properties of our
data set. Our method for measuring the shear and our treatment of
systematic effects are described in \S\ref{procedure}. Our results are
presented in \S\ref{results} and discussed and summarized in
\S\ref{conclusions}.

\section{Theory}
\label{theory}
We measure the average shear $\overline{\gamma}_{i}$
within each WFPC2 chip, which are square cells of side
$\alpha=1.27'$. The angular fluctuations of the resulting smoothed
shear field are characterized by the shear correlation functions
$C_{i}(\theta) \equiv \langle \overline{\gamma}_{i}^{r}(0)
\overline{\gamma}_{i}^{r}(\theta) \rangle$, where $i=1,2$ and the
brackets refer to an average over pair of cells separated by an angle
$\theta=|{\mathbf \theta}|$. The shear components $\gamma_{i}^{r}$ are
measured in a rotated coordinate system whose $x$-axis is aligned with
the separation vector ${\mathbf \theta}$ between the two points (see e.g.
Heavens, Refregier \& Heymans 2000). These correlation functions can be computed for any
cosmological model using (see eg. BRE)
\begin{equation}
\label{eq:ci_th}
C_{i}(\theta) = \frac{1}{4\pi} \int_{0}^{\infty}~lC_{l}
\left| \widetilde{W}_{l} \right|^{2} \left[ J_{0}(l\theta)
\pm J_{4}(l\theta) \right], ~~i=1,2,
\end{equation}
where $\widetilde{W}_{l}$ is the window function of a square cell.  The
shear power spectrum $C_{l}$ is defined as in BRE and can be 
evaluated from the evolution of the non-linear
matter power spectrum using the fitting formula from Peacock \& Dodds
(1997). 
In this paper, we will consider a $\Lambda$CDM model with
$\Omega_{m}=0.3$, $\Omega_{\Lambda}=0.7$, $\Gamma=0.25$, and consider
several values of $\sigma_{8}$, the normalization of the mass
fluctuations on 8 $h^{-1}$ Mpc scales.

The variance of the cell-averaged shear is  $\sigma_{\rm lens}^{2} \equiv
\langle |\overline{\gamma}|^{2} \rangle = C_{1}(0)+C_{2}(0)$.  For the
$\Lambda$CDM model, the  rms shear in a cell is
\begin{equation}
\label{eq:sigma_lens}
\sigma_{\rm lens} \simeq 0.0197 
 \left( \frac{\sigma_{8}}{1} \right)^{1.24} 
 \left( \frac{\Omega_{m}}{0.3} \right)^{0.60}
 \left( \frac{z_{m}}{0.9} \right)^{0.85},
\end{equation}
where $z_{m}$ is the median redshift of the background galaxies, whose
redshift distribution was taken to be of the form of
Equation~(\ref{eq:red_dist}) below. The scaling relations in
Equation~(\ref{eq:sigma_lens}) provide an
excellent approximation in our range of interest, namely, $0.6 \la
\sigma_{8} \la 1.4$ and $0.7 \la z_{m} \la 1.2$.


\section{Data}
\label{data}
The Hubble Space Telescope Survey Strip (sometimes referred to as the
`Groth Strip', or the `Groth-Westphal Strip') was taken during March
and April 1994 using the Guaranteed Time Observations of one of us
(EJG) and the WFPC-1 Instrumentation Definition Team (Groth et
al. 1994).  The Strip was observed in two passbands, V (F606W) and I
(F814W).  For 27 of the pointings in the Strip, total exposure times
were 2800 and 4400 seconds in the V and I bands, respectively.  One of
the Strip pointings (the Strip Deep Field) had exposure times of
24,400 and 25,200 seconds, arranged in 4 dithers.  We used only the
I-band images to study weak lensing. The Strip images consist of 4
individual exposures with equal exposure times. Here, we chose to use
only one of the four dithers for the Deep Field, thus avoiding the
complications in the PSF correction that dithering would introduce
(see \markcite{rho99} RRG).

We used the Faint Object Classification and Analysis System (FOCAS;
Jarvis and Tyson, 1981) within IRAF to create a catalog of objects in
the Strip.  We determine the sky background and its standard deviation
using the IRAF task {\tt Imarith}. This was found to be more reliable
than the corresponding algorithm within FOCAS. Magnitude zero points
are those defined by \markcite{hol95} Holtzman et al. (1995). A FOCAS
catalog was created for each filter, and a matched catalog was created
for objects that had I and V positions that coincided within a radius
of about 5 pixels.  There were approximately 10,000 matched objects
with I$<26$.  Of these objects, approximately 4000 were classified as
galaxies, had well defined moments and were sufficiently large to be
used in our weak lensing study. We were careful not to double count
galaxies that fall in the small overlap region between consecutive
fields in the Strip.

Koo et al. (1996)\markcite{koo96} have shown that a
small subsample (24 galaxies) of  Strip objects matches an
extrapolation of the CFRS sample redshift distribution. 
More recent redshift measurements of galaxies in the Strip by
the DEEP collaboration (DEEP Collaboration, 1999) agree with the preliminary
findings in Koo et al. (1996).\markcite{koo96}
Assuming the functional form of the redshift distribution found
in the CFRS, 
\begin{equation}
\label{eq:red_dist}
\frac{dN}{dz} \propto z^2 e^{-(z/z_{o})^2}
\end{equation}
and using the median redshift of $z_{m} \simeq z_{o}=0.5$ for objects in the
range $17<I<22$ found in that survey \markcite{lil95}(Lilly et
al. 1995), we calculate a median redshift for objects in the 
Strip.  The galaxy sample we use to study weak lensing ($I<26$ and
size $d>1.5$ pixels) has a median magnitude of $I=23.6$.
Extrapolating from the CFRS sample using the  Strip number
counts, this corresponds to a median redshift of $z_{m}=0.9$ (Rhodes
1999, Groth and Rhodes, 2001)\markcite{rho99}\markcite{gro00}.
Photometric redshifts for galaxies in the Hubble Deep Fields
(HDF-North and HDF-South) have been measured by Lanzetta et
al. (1996). A linear fit to the median redshift versus median
magnitude plot yields $z_m=0.87$ for the HDF-South and $z_m=1.00$ for
the HDF-North for a median magnitude of $I=23.6$, where we have used the
magnitude conversion $I=-0.49+I_{AB}$. This is consistent with what we
found above using an extrapolation of the CFRS sample. Therefore,
we adopt a median redshift of $z_m=0.9\pm 0.1$ for the lensing
analysis in this paper.

\section{Procedure and Systematic Effects}
\label{procedure}
The procedure we use for extracting galaxy ellipticities and shear
from the source images is described in detail in RRG\markcite{rho99}
(see also \markcite{rhot99}Rhodes 1999). This method is based on one
introduced by Kaiser, Squires, and Broadhurst (KSB, 1995), but
modified and tested for application to HST images.
The KSB method was shown to be adequate for current ground based
surveys (\markcite{bac00}Bacon, et al. 2000b; \markcite{erb00}Erben et
al. 2000). In RRG, we used numerical simulations and an array of tests
to show that our method was  sufficiently accurate for WFPC2
images.

We use our method to correct for two effects: camera distortion and
convolution with the anisotropic PSF. Corrections are done using
moments measured with a Gaussian weight function whose size in pixels,
$\omega$, depends on the size of the object as $\omega=\max(2,
\sqrt{A/\pi})$, where $A$ is the area of the object in pixels.  The
minimum size of two pixels was found to be the optimal weight function
size for stellar objects using both actual WFPC2 data and simulations
using artificial stellar images created using the program Tiny Tim
(Krist and Hook 1997).

Camera distortions are corrected using a map derived
from stellar astrometric shifts (Holtzman, et al., 1995). 
PSF corrections are determined from HST observations of two
globular clusters (M4 and NGC6572).
Finally, we
measure the ellipticities ($\epsilon_{i}$) of the galaxies in the Strip and convert them
into shear estimates using $\gamma_{i}=G^{-1} \epsilon_{i}$, where
$G$ is the shear susceptibility factor given by Eq.~(30) in RRG.

RRG have shown that the residual systematic effects can be
greatly minimized if (1) the shear is averaged over entire 
 chips, and (2) only ``large'' galaxies with a (convolved) 
rms radius greater than 1.5 pixels (or about 0.15'') are
selected. Since the PSF anisotropy is mainly tangential about the chip
centers, condition (1) ensures that the mean PSF ellipticity is small
($\epsilon^{*} \sim 0.02$). Since the shapes of larger galaxies are
less affected by the PSF, condition (2) ensures that the impact of the
PSF anisotropy on galaxy ellipticities $\epsilon^{g}$ is small.
Specifically, RRG show that the ellipticity of the selected galaxies
$\epsilon_{g}$ induced by the PSF ellipticity is reduced by a factor
$f_{\rm red} \equiv \epsilon^{g}/\epsilon^{*} \simeq 0.13$.

As shown in RRG, the residual systematics for this galaxy subsample
are dominated by two effects.  The first arises from the imperfect
accuracy of the ellipticity corrections and shear measurement
method. This effect results in an uncertainty in the normalization of
the shear of about 4\% and is easily incorporated at the end of our
analysis. The second results from the time variability of the PSF. To
quantify this effect, we measure the mean ellipticity of the PSF in
each chip for 4 stellar fields (M4 observed at 2 different epochs,
NGC6572, and a stellar field from the WFPC2 parallel archive).  We
find that the chip-averaged PSF ellipticity
$\overline{\epsilon}^{*}_{ic}$ varies by about 0.01 (rms) in chips
$c=2$ and 3 and 0.02 in $c=4$.  The variations appear to be stochastic
and to be unrelated to the focus position of the telescope. These time
variations must be corrected statistically since the small number of
stars in the Strip precludes measuring the PSF in each field
individually. For this purpose, we measure the covariance
$C^{*}_{ic,jc'} \equiv {\rm
cov}[\overline{\epsilon^{*}_{ic}},\overline{\epsilon^{*}_{jc'}}]$ of
the stellar ellipticities by averaging over all stellar fields, and
convert this into the systematics shear correlation matrix $C^{\rm
sys}_{ic,jc'} \simeq f_{\rm red}^{-2} G^{-2} C^{*}_{ic,jc'}$  which
is subtracted from the galaxy shear correlation functions. The
diagonal elements are the shear variance produced by the time
variability of the PSF $\sigma_{\rm sys}^{2}=\langle C^{\rm
sys}_{1c,1c} + C^{\rm sys}_{2c,2c} \rangle_{c} \simeq 0.0011^{2}$,
where the average is over all chips.

\section{Results}
\label{results}
To study the lensing statistics in the Strip, we first compute the
mean shear $\overline{\gamma}_{ic}$ in each chip $c$, by averaging
over all selected galaxies in the chip. This mean shear is the sum of
contributions from lensing, noise and systematic effects,
i.e. $\overline{\gamma}_{ic}=\overline{\gamma}^{\rm lens}_{ic}
+\overline{\gamma}^{\rm noise}_{ic}+\overline{\gamma}^{\rm sys}_{ic}$.
From the distributions of the galaxy shears within each chip, we can
measure the covariance matrix $C^{\rm noise}_{ic,jc'} \equiv {\rm
cov}[\overline{\gamma}^{\rm noise}_{ic}, \overline{\gamma}^{\rm
noise}_{jc'}]$ of the mean shear components. Since the chips do not
overlap, $C^{\rm noise}_{ic,jc'}$ vanishes for $c\neq c'$. The
diagonal elements $C^{\rm noise}_{ic,ic} \equiv \sigma_{{\rm
noise},ic}^{2}$ are the errors in the mean for the mean shear
components $\overline{\gamma}_{ic}$. The average noise variance is
$\sigma_{\rm noise}^{2} \equiv \langle \sigma_{\rm noise,1c}^{2} +
\sigma_{\rm noise,2c}^{2} \rangle_{c} \simeq 0.047^{2}$, where the
average is over all chips. The off-diagonal entries in the noise
matrix are an order of magnitude below the diagonal ones and thus do
not have much impact on the analysis. We nevertheless keep them for
completeness.

The  shear pattern for the Strip is shown in
Figure~\ref{fig:Strip_ellip}. Since the expected signal on the chip
scale is about $\sigma_{\rm lens} \simeq 0.02$ (see Eq.[2]), the
signal-to-noise ratio $\sigma_{\rm lens}/\sigma_{\rm noise}$ for each
chip is about 0.4. The lensing signal must therefore be searched for
statistically. For this purpose, we first consider the mean shear
$\langle \overline{\gamma_{ic}} \rangle_{c}$ averaged over all chips
in the Strip. We find $\langle \overline{\gamma_{1}} \rangle = 0.005
\pm 0.004$ and $\langle \overline{\gamma_{2}} \rangle = 0.005 \pm
0.004$, where the errors are $1\sigma$ errors in the mean. This is
consistent with zero, as expected given the length of the Strip. This
nonetheless confirms that we are not subject to systematic effects on
large scales.

\vspace{0.4cm}
\centerline{{\vbox{\epsfxsize=3.7truein\epsfbox{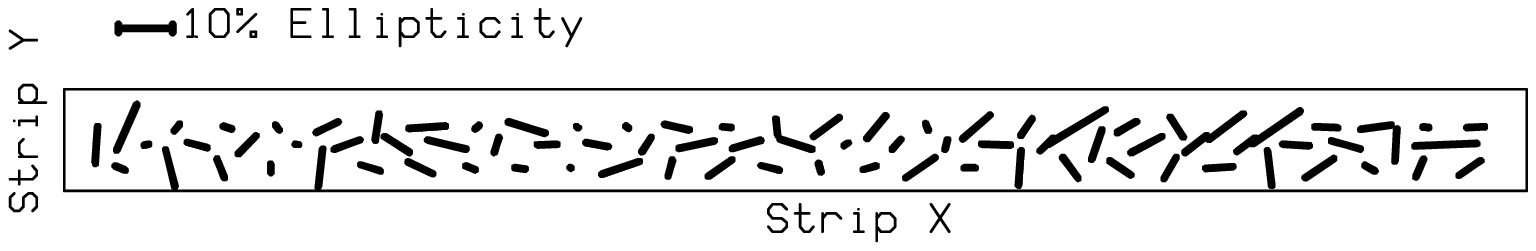}}}}
\vskip-0.3cm
\figcaption{\small The ellipticity of each field and chip in the Survey
Strip.  The Strip measures about 4 by 42 arcminutes.
\label{fig:Strip_ellip}}
\vspace{0.4cm}

We next consider the variance of the shear on the chip scale. In
general, the total observed shear variance $\sigma_{\rm tot}^{2}
\equiv \langle |\overline{\gamma}_{ic}|^{2} \rangle$ is the sum of the
contribution from lensing, from noise, and from systematic effects,
i.e. $\sigma_{\rm tot}^{2} = \sigma_{\rm lens}^{2} + \sigma_{\rm
noise}^{2} + \sigma_{\rm sys}^{2}$. An estimator for the lensing
variance is thus $\sigma_{\rm lens}^{2} \simeq \sigma_{\rm tot}^{2} -
\sigma_{\rm noise}^{2} - \sigma_{\rm sys}^{2}$ (for more details, see
BRE). Using our estimate of $\sigma_{\rm sys}$ (see
\S\ref{procedure}), we find $\sigma_{\rm lens} \simeq 0.018$, which is
close to the expected value of 0.0197 in the $\Lambda$CDM model (see
Eq.~[\ref{eq:sigma_lens}]). Estimating errors on the shear variance is
complicated since the chips are contiguous and thus not
independent. Instead of attempting to do this, we use a maximum
likelihood approach to extract the lensing signal on all scales
simultaneously.

To do so, we assume that the shear field obeys Gaussian
statistics. Since we are averaging over a large number of galaxies
(about 50) in each field, the central value theorem ensures that this
assumption is valid for the noise contribution. It is difficult to
establish whether the systematics are Gaussian; however, since they
have a much smaller amplitude than the noise, it is sufficient to
assume that they are so. The lensing shear field is known not to be
Gaussian, especially on scales smaller than 10' which are dominated by
the non-linear evolution of structures (\markcite{jai97}Jain \& Seljak
1997). However, as we have shown above, the noise dominates over the
lensing signal in each individual chip. Therefore our Gaussian approximation should not 
underestimate our error bars by a large amount.

With this approximation, the probability of measuring the mean shears
$\overline{\gamma}_{ic}$ given a cosmological model $M$ is
\begin{equation}
P(\overline{\gamma}_{ic}|M) \simeq \frac{e^{-\frac{1}{2} g^{T} C^{-1}
g}}{(2\pi)^{N}|C|^{\frac{1}{2}}}
\end{equation}
where $N=84$ is the number of chips, and
$g=(\overline{\gamma}_{11},\overline{\gamma}_{21},\cdot \cdot \cdot ,
\overline{\gamma}_{1N}, \overline{\gamma}_{2N})$ is the data vector
composed of the $2N$ measured shears. The total correlation matrix
$C$ is given by $C=C^{\rm lens}+C^{\rm noise}+C^{\rm sys}$.  The
systematics correlation matrix $C^{\rm sys}$ is computed as explained
in \S\ref{procedure}, assuming that each field is independent as
indicated by the stochastic nature of the PSF variations. The lensing
correlation matrix $C^{\rm lens}$ depends on the cosmological model
and can be computed by unrotating the rotated correlation functions
$C_{1}(\theta)$ and $C_{2}(\theta)$ (Eq.~\ref{eq:ci_th}). Following
Bayes' postulate, the likelihood $L$ of the model $M$ given the data
is then $L \propto P(\overline{\gamma}_{ic}|M)$, so that its logarithm
is given by
\begin{equation}
\label{eq:lnl}
- 2 \ln L \simeq g^{T} C^{-1} g + \ln |C| + {\rm
constant}.
\end{equation}

First, we test the null hypothesis, namely the absence of lensing
corresponding to $C^{\rm lens}=0$. In this case, $C=C^{\rm
noise}+C^{\rm sys}$ is a constant, and the log likelihood reduces to
$- 2 \ln L = \chi^{2} + {\rm constant}$, where $\chi^{2} \equiv g^{T}
C^{-1} g$. For the Strip, we find a reduced $\chi^{2}$ of 1.31 with
$2\times 84$ degrees of freedom. The null hypothesis is thus ruled out
at the 99.5\% confidence level, thus establishing our detection of the
lensing signal. Note that this null test does does not rely on the
assumption that the lensing shear field is Gaussian, because we have set it to 
zero in this case.

Now that we have established the presence of a lensing signal, we can
compute the constraints our data impose on cosmological parameters.
To do so, we compute the Maximum Likelihood (ML) estimator for
$\sigma_{8} (\Omega_{m}/0.3)^{0.48}$. This was done by maximizing $\ln
L$ (Eq.~[\ref{eq:lnl}]), which was computed with $C_{\rm lens}$
matrices corresponding to different values of this parameter. To
compute the error in $\sigma_{8}$ we performed 1000 random
realizations of the Strip with shears drawn from a multivariate
Gaussian distribution with a covariance matrix given by $C=C^{\rm
lens}+C^{\rm noise}+C^{\rm sys}$ (with $C^{\rm lens}$ at the MLE value
of $\sigma_{8}(\Omega_{m}/0.3)^{0.48}$ of the Strip).

\centerline{{\vbox{\epsfxsize=3.2truein\epsfbox{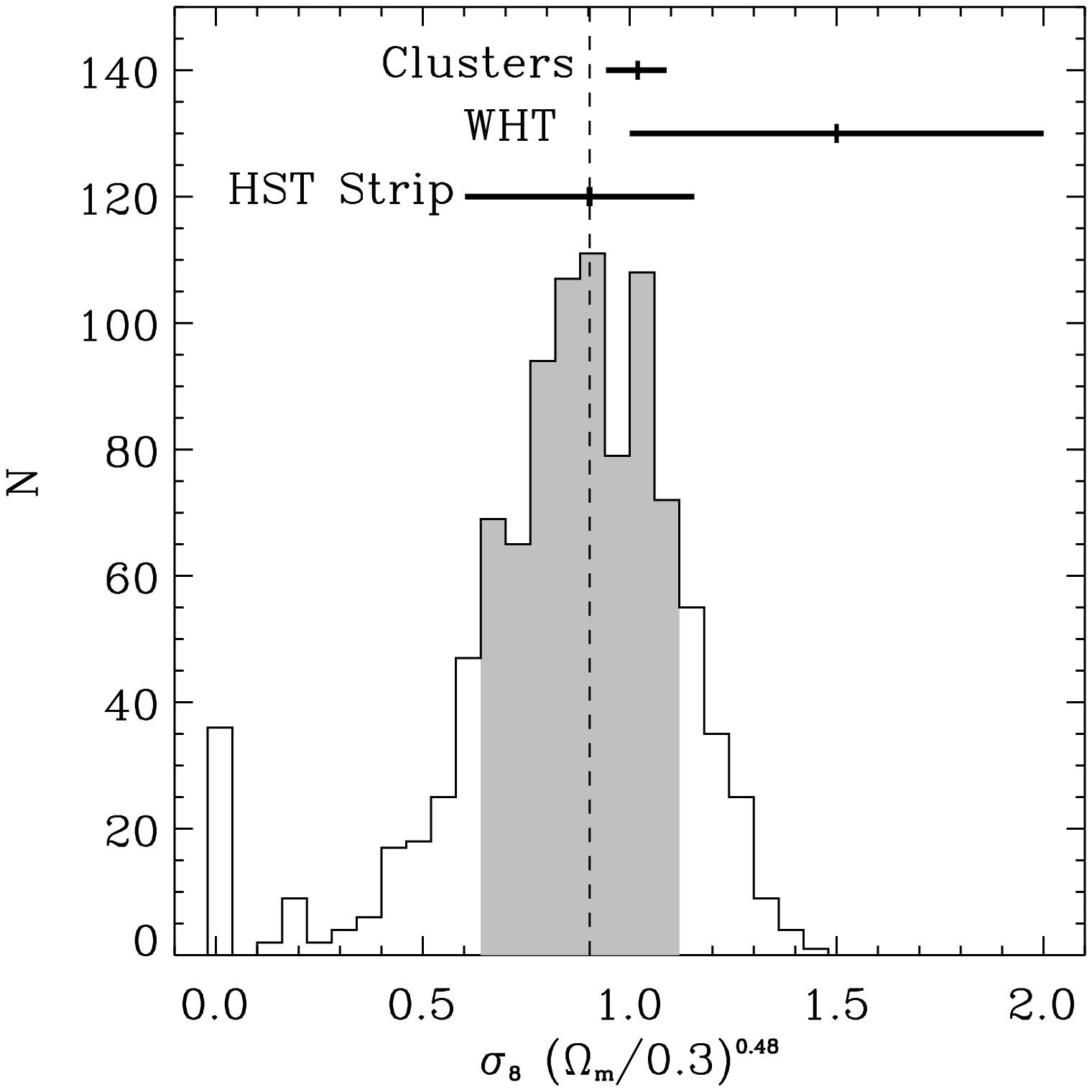}}}}
\figcaption{\small Distribution of the ML normalization
of the power spectrum $\sigma_{8} \Omega_{m}^{0.48}$ derived from 1000 random
realizations of the Strip. The normalization corresponding to the
real Strip is shown as the vertical dashed line. The 68\% confidence
level is shown as the shaded region. The full 68\% confidence
interval which includes (Gaussian) cosmic variance, the uncertainty
in the median redshifts of the background galaxies and that arising
from systematic effects is shown as a horizontal bar. It can be
compared to the corresponding limits derived from the William Herschel
Telescope (WHT) cosmic shear survey of BRE,
and to that from cluster abundance (Pierpaoli et al. 2000), shown as
marked horizontal bars.
\label{fig:sigma8}}
\vspace{3mm}

The resulting
distribution of the ML values is shown in Figure~\ref{fig:sigma8}.
We thus obtain $\sigma_{8} (\Omega_{m}/0.3)^{0.48} =
0.90^{+0.22}_{-0.24}$ where the error bars correspond to the 68\%
confidence interval. After propagating errors induced by the uncertain
median redshift $z_{m}$ (see \S\ref{data}) of the galaxies and by the
shear measurement method, our final 68\% confidence range becomes
$\sigma_{8} (\Omega_{m}/0.3)^{0.48} = 0.90^{+0.25}_{-0.30}$ and is
shown as a horizontal bar in Figure~\ref{fig:sigma8}.  This value is
consistent with the William Herschel Telescope (WHT) cosmic shear
survey of BRE, who found $\sigma_{8} (\Omega_{m}/0.3)^{0.48} = 1.5 \pm
0.5$, as shown by the associated horizontal bar on the Figure. Our
results are also consistent with the amplitude of the cosmic shear
signals found by other groups (Wittman et al. 2000; van Waerbeke et
al. 2000; Kaiser et al. 2000; Maoli et al. 2000). However, since these
groups do not include cosmic variance and redshift uncertainty in
their error bar estimates, their results can not be directly compared
with ours.  The normalization from cluster abundance $\sigma_{8}
(\Omega_{m}/0.3)^{0.60} = 1.019^{+0.070}_{-0.076}$ (Pierpaoli et
al. 2000) is consistent with ours and is also shown on the figure
(after ignoring the small difference in the exponents for
$\Omega_{m}$).


\section{Conclusions}
\label{conclusions}
We have detected, for the first time, cosmic shear using space-based
images. Our detection in the HST Survey Strip is significant at the
99.5\% confidence level. Using a ML method to search for a lensing
signal on all scales simultaneously, we derived a normalization of the
matter power spectrum of $\sigma_{8} \Omega_{m}^{0.48} =
0.51^{+0.14}_{-0.17}$. This 68\% confidence interval includes
statistical noise, (Gaussian) cosmic variance, uncertainty in the
redshift distribution of the galaxies and in our shear measurement
method. This is consistent with ground-based cosmic shear surveys and
with the normalization derived from cluster abundance.

Our result demonstrates the power of space-based images to measure
cosmic shear. The small PSF in HST images affords both a larger
surface density of background galaxies and a much reduced impact of
the systematics induced by the PSF anisotropy. The latter advantage
will be even more important for future, wider surveys whose sensitivity
will be comparable to the amplitude of the systematics.

Our current results are chiefly limited by the relatively small area
of the  Strip. The signal-to-noise ratio for the detection of
cosmic shear  in $N_{f}$ independent WFPC2 fields, in which we cannot search for lensing on all scales
simultaneously as we have done here, is
approximately S/N $\simeq (\sigma_{\rm lens}/\sigma_{\rm noise})^{2}
(3N_{f})^{\frac{1}{2}} \simeq 6.8 (N_{f}/500)^{\frac{1}{2}}$ (see BRE).
Our detection can thus be significantly improved in the future by an
analysis of existing disjoint WFPC2 fields.

\acknowledgments We thank David Bacon for useful discussions.  JR was supported by NASA Grant NAG5-6279 and
a National Research Council-GSFC Research Associateship. AR was supported by a EEC fellowship from the TMR
Gravitational Lensing Network and by a Wolfson College Research
Fellowship. EJG was supported by NASA Grant NAG5-6279. We thank the WFPC1 IDT for  their cooperation.

\newpage

\end{document}